# PERFORMANCES OF A NEWLY HIGH SENSITIVE TRILAYER F/CU/F GMI SENSOR

F. Alves[1], B. Kaviraj[1], L. Abi Rached[1], J. Moutoussamy[2] and C. Coillot[2]

[1]LGEP/SPEE Labs; CNRS UMR8507; Supélec; Univ Pierre et Marie Curie-P6; Univ Paris Sud-P11,
11 rue Joliot-Curie, Plateau du Moulon, 91192 Gif sur Yvette, FRANCE
(Tel :+33169851657, E-mail: francisco.alves@lgep.supelec.fr, bhaskar.kaviraj@lgep.supelec.fr, lena.abirached@lgep.supelec.fr)

[2] CETP/IPSL/CNRS UMR 8639, 10-12, Avenue de l'Europe, 78140 Vélizy, FRANCE
(Tel : +33139254858, E-mail: christophe.coillot@cetp.ipsl.fr, jmo@cetp.ipsl.fr)

**Abstract:** We have selected stress-annealed nanocrystalline Fe-based ribbons for ferromagnetic/copper/ferromagnetic sensors exhibiting high magneto-impedance ratio. Longitudinal magneto-impedance $\Delta Z/Z_{sat}$ reaches 400% at 60 kHz and longitudinal magneto-resistance $\Delta R/R_{sat}$ increases up to 1300% around 200 kHz.

**Keywords:** Magneto-impedance, stress-annealing, nanocrystalline materials, magnetic sensors

## 1. Introduction

Detection of embedded cracks in non destructive inspection, study of solar wind interactions with terrestrial magnetosphere, molecular recognition systems and selective detection require high sensitive and small sized magnetic sensors. Magneto-impedance (MI) in sandwich ferromagnetic/conductor/ferromagnetic films is being actively investigated nowadays to miniaturise MI elements and maintain its high sensitivity for micro-sensing applications. The principle of GMI effect consists of a significant change of the impedance value of a magnetic conductor (wire, ribbon, thin layers) crossed by a high frequency current $I_{ac}$, when it is subjected to quasi-dc or low frequency ac magnetic field B [1,2]. The applied magnetic field affects the transversal permeability of the magnetic conductor, determining modifications of the AC current penetration depth, which are closely related to the impedance value at a given frequency [3,4]. The GMI effect has been investigated in a variety of Fe- and Co-based wires, ribbons and films. A very high sensitivity to an external field is typical of magneto-impedance (MI) in soft ferromagnetic conductors with well-defined anisotropy [5,6]. Magnetic sensors based on MI in amorphous ferromagnetic conductors have been developed which demonstrate the field detection resolution of $10^{-6}$ Oe ($10^{-7}$ mT) for the full scale of $\pm 1.5$-2Oe (0.15-0.2mT) with a sensor head length of 1mm [7]. A very sensitive MI has been reported to occur in F/C/F multilayers, in which the impedance change ratio is several times larger than in a similar single layer ferromagnetic film. For example, in CoSiB/Cu/CoSiB films of 7μm thick, the MI ratio is 340% for a frequency of 10MHz and a dc magnetic field of 9 Oe (0.9 mT) [8]. On the other hand, in the CoSiB layer of the same thickness, the impedance varies over few percents under these conditions. In electrically uniform materials under the condition of a strong skin effect, the impedance is a square root function of the frequency and permeability but in the multi-layers, a very large change in impedance can be observed at much lower frequencies when the inductance related to the outer magnetic layers is much higher than the resistance determined mainly by the inner conductor [9]. Then, the impedance varies linearly with the frequency and permeability and due to this advantage, MI in sandwich films has a potential to be used in developing sensitive micro-magnetic sensors and magnetic heads for high density magnetic recording.

In this paper, we present the latest developments in the research of trilayer ferromagnetic/conductor/ferromagnetic (F/C/F) giant magneto-impedance (GMI) sensor dedicated to these applications. The GMI ratio is calculated from impedance-field Z(B) curve as:

$$\Delta Z/Z_{s\,at}\,(\%) = 100 \times (Z(B) - Z(B_{sat}))/Z(B_{sat}) \quad (1)$$

where $B_{sat}$= 8 mT.

The measurements of dc/ac magnetic fields have been carried out using trilayer F/C/F nanocrystalline GMI sensor.

## 2. Measurements of DC fields

$Fe_{75}Si_{15}B_6Cu_1Nb_3$ ribbons, 20 μm thick, were prepared by planar flow casting technique and purchased by Imphy Alloys. After short annealing under stress, nanocrystalline $Fe_{75}Si_{15}B_6Cu_1Nb_3$ ribbons exhibit extraordinary linear and non-hysteretic characteristics with controlled transversal anisotropy field, $H_k$ (for details, see ref. [10]), extreme low temperature dependence (0.17%/°C). Fig. 1 shows the magnetic domain patterns of the stress-annealed ribbons obtained using the surface magneto-optical Kerr effect (SMOKE) technique. As the stress is increased during treatment, magnetic structures are refined. Furthermore, such annealed ribbons present the particularity to be ductile and are easily handled for the elaboration of F/C/F glued layered structure for GMI sensor. The layers of ribbons are 20μm thick, 6cm long and 1cm wide. Conductor/magnetic ribbon width ratio is 5:10.

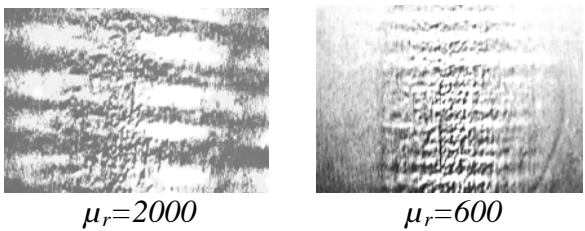

$\mu_r=2000$     $\mu_r=600$

Fig. 1.  1mmx2mm SMOKE pictures.

By means of two pairs of Helmholtz coils providing the dc field $H_{dc}$, two configurations have been tested: (i) $H_k$ perpendicular to ribbon axis and $H_{dc}$ parallel to $I_{ac}$ (Longitudinal Magneto-Impedance; LMI), (ii) $H_k$ and $H_{dc}$ perpendicular to $I_{ac}$ (Transversal Magneto-Impedance; TMI). Impedance measurements were performed within 30-600 kHz range using a SR844 (Stanford Research Systems) Lock-in Amplifier with input filters 6dB/oct. A 10Ω, 0.1% resistor is used to set the current $I_{ac}$ at a value between 10 and 20$mA_{RMS}$.

Fig. 2 confirms two well-known theoretical results in trilayer F/C/F structures: (i) the impedance variation is several times larger than in a similar single layer ferromagnetic film and enhanced by the conductivity of the conductor, (ii) in the same time, maximum LMI occurs at smaller frequency. Fig.3 shows a typical GMI curve requiring a bias field to achieve optimal sensitivity-to-field. Table 1 gives some values of the maximum sensitivity-to-field obtained using real ($S_{/R}$) or imaginary ($S_{/X}$) parts of impedance. As comparison, commercial GMR sensors exhibit $S_{/R}$ =0.35%/mT [11].

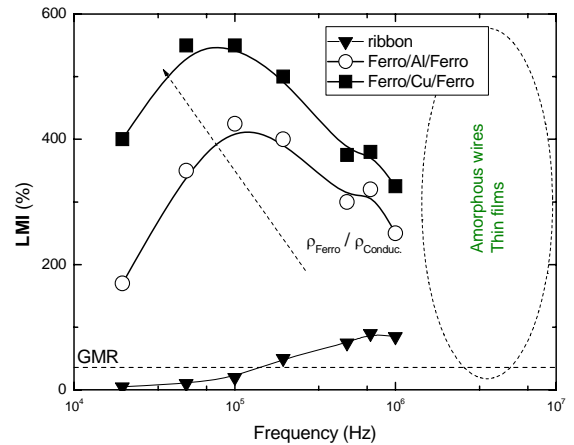

Fig. 2. Relative variation of GMI voltage versus frequency for different F/Conductor/F structure.

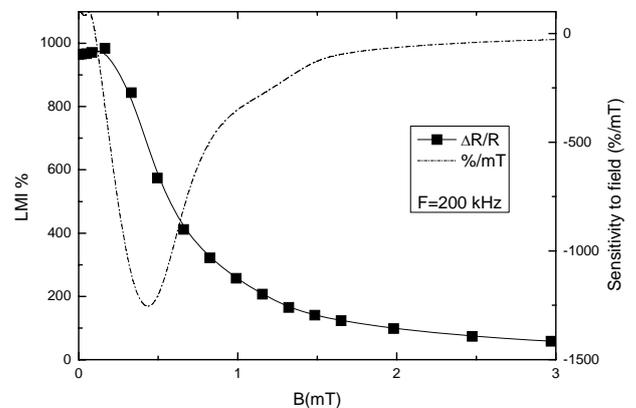

Fig. 3. Relative variation of real part of GMI voltage and sensitivity-to-field along F/Cu/F sensor.

Table 1 Maximum absolute sensitivities-to-field of F/Cu/F GMI sensor used in the LMI configuration.

| F(kHz) | 30 | 60 | 100 | 200 | 600 |
|---|---|---|---|---|---|
| $S_{/R}$(%/mT) | 175 | 430 | 770 | 1300 | 1200 |
| $S_{/X}$(%/mT) | 610 | 500 | 470 | 385 | 150 |

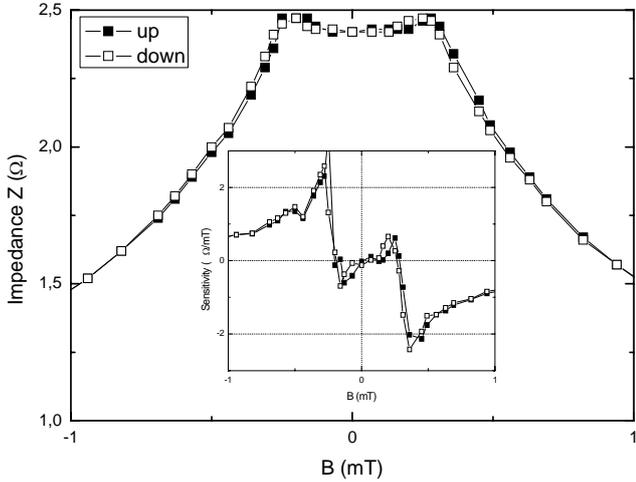

Fig. 4. Impedance versus external field. In inset, the derivative of impedance (sensitivity). The value and frequency of the ac driving current are respectively 3 mA and 600 kHz.

Fig.4 presents the dependence of impedance and sensitivity (Ω/mT) with field applied along the trilayer F/Cu/F structure which has been submitted to an increase from $-B_{sat}$ to $B_{sat}$ and then decrease to $-B_{sat}$. We observed very low hysteresis, anisotropy field around 200 A/m, maximum sensitivity of 2 Ω/mT.

## 3. Measurement of low AC fields

Especially designed electronic amplifier has been used for the F/Cu/F GMI biased sensor. The current excitation through the GMI acts as a simple "amplitude modulation". Thus, the magnetic field measurement is contained inside the GMI signal. Then a ferrite transformer together with a low noise differential amplifier permits the GMI signal amplification with very low added noise. Finally, the output signal is multiplied by clock (in phase with the excitation signal) in order to perform a synchronized extraction of the measured magnetic field. Fig 5a illustrates a typical spectrum noise density measurement. Here, the sensor is excited at 600 kHz (HF peak) and detects an external ac field of 65 Hz (LF peak). By changing the bias current, we can follow the LF/HF peaks which are characteristics of respectively transfer function (fig. 5b) and sensitivity (fig. 5c). The results obtained with spectrum noise analyzer are well consistent with the previous dc field measurements made with lock-in amplifier.

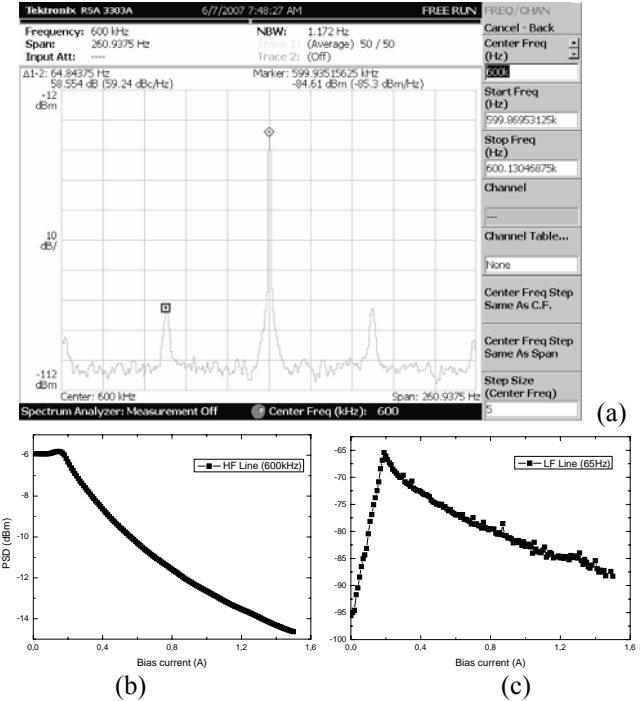

Fig. 5. Spectrum analyzer measurement (a), transfer function (b) and sensitivity (c) evolutions with bias current.

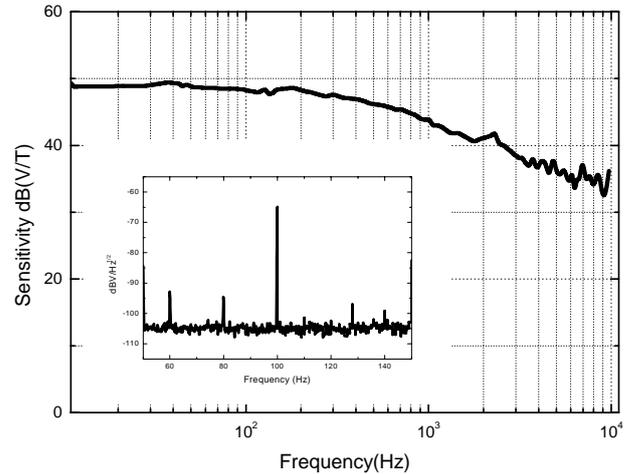

Fig. 6. Sensitivity versus frequency. In inset, spectrum noise density of a GMI sensor under ac field of low frequency (100 Hz).

Fig 6 shows the frequency dependence of sensitivity: 60V/T at the GMI level or 4500V/A/T (to be compared to 750V/A/T for

AlGaAs/InGaAs/GaAl best Hall sensors working with nominal current of 1 mA [12]) up to a few hundred Hz. We have also checked the high sensitivity of F/Cu/F GMI sensor: 1m diameter Helmholtz coils are used to generate an AC field (100Hz) of 500 nT. In these conditions the spectrum noise density (PSD) of demodulated signal (Vs1) exhibits a peak at 100Hz (cf. fig.6) with a magnitude of -63.6 dBV.Hz$^{-1/2}$ (spectrum analyzer frequency range 50-150Hz, span 1600, thus ΔF=62.5 mHz). Consequently, the amplitude of 100Hz line is 165μV when the applied magnetic field reaches 500nT. The Noise Equivalent Magnetic Induction (NEMI) is equal to:
-105 dBV.Hz$^{-1/2}$ -51.6 dBV/T = -156.6 dBT.Hz$^{-1/2}$ or 13nT/Hz$^{-1/2}$.

An important point is that the background noise level of the GMI sensor is not reached as the noise from oscillator is preponderant. Further works will aim to investigate added noise from oscillator.

## 4. Conclusions

First results obtained in F/Cu/F GMI biased sensors using nanostructured FeSiBCuNb materials concern sensitivity to DC field (thousand %/mT), sensitivity to AC field (60 V/T up to a few hundred Hz), no hysteresis is observed. At the current stage of development, NEMI performances are limited to electronic noise, which should be overcame.

## Acknowledgement

This work is partially supported by a grant from the National Research Agency (number ANR-05-NANO-067-04).

## Biographies

**Francisco Alves** is professor at University of Paris-Sud (France) since 2005. He received his PhD degree in electrical engineering in 1993 and his "Habilitation à Diriger des Recherches" in 2002. In 2005, he joined the Electrical Engineering Laboratory (LGEP), Gif-sur-Yvette, France. His research interests include characterization of nanocrystalline alloys, stress-annealing treatments (patent) and fabrication of AMR or MI (patent) sensor arrays.

**Bhaskar Kaviraj** is currently in post-doctoral position **in** the Electrical Engineering Laboratory (LGEP), Gif-sur-Yvette, France. His research interests include fabrication and characterization of sandwiched sputtered Ferromagnetic /copper/ferromagnetic sensors based on giant magneto-impedance effect.

**Léna Abi Rached** is a PhD student at the University of Paris-Sud (France). In 2006, she joined LGEP laboratory to pursue her PhD degree in high sensitive MI sensors dedicated to biomedical applications.

**Joel Moutoussamy** was born in Basse-Terre, Guadeloupe, France, in 1970. He is a full time teacher since 1996 at Dijon technological high school, G. Eiffel. He teaches object-oriented programming, instrumentation and real-time systems in undergraduate technical classes (BTS). In 2003, he joined the CETP laboratory (Center for the Study of Earth and Planet Environments), Velizy, France, to pursue his PhD degree in wide bandwidth magnetometers for space-oriented applications.

**Christophe Coillot** was born in Choisy le Roi, France, in 1971. He received the PhD degree in electronics from the University of Montpellier, France, in 1999. In 2001, he joined the CETP as research engineer to manage magnetometer instrument design for space applications and to participate in the research activity of CETP in magnetometry. He is currently involved in some spacecraft experiments, the two famous one are: Bepicolombo for Mercury environment and THEMIS for Earth environment. His field of interest concerns low noise and low power consumption preamplifier, search coil improvement through modelling and optimization and more recently: AMR and Hall effect sensors.